\documentclass[aps,prl,reprint,preprintnumbers]{revtex4-1}
\usepackage{graphicx}
\usepackage{bm}
\usepackage{epsfig}
\usepackage{ulem}
\usepackage{color}
\usepackage{mathrsfs}
\usepackage{dcolumn}
\usepackage{setspace}
\usepackage{array}
\usepackage{amsmath}
\usepackage{amssymb}
\usepackage{dsfont}
\usepackage{gensymb}
\usepackage{dcolumn}
\usepackage{multirow}
\usepackage{bibentry,natbib}
\usepackage{booktabs}
\begin{document}
\bibliographystyle{unsrt}
\title{Quantum Atmospherics for Materials Diagnosis}

\author{Qing-Dong Jiang$^1$, Frank Wilczek${^1}{^2}{^3}{^4}$}
\affiliation{{}\\ $^1$Department of Physics, Stockholm University, Stockholm SE-106 91 Sweden\\
$^2$Center for Theoretical Physics, Massachusetts Institute of Technology, Cambridge, Massachusetts 02139 USA\\
$^3$T. D. Lee Institute, and Wilczek Quantum Center, Department of Physics and Astronomy, Shanghai Jiao Tong University, Shanghai 200240, China\\
$^4$Department of Physics and Origins Project, Arizona State University, Tempe AZ 25287 USA}
\begin{abstract}
Symmetry breaking states of matter can transmit symmetry breaking to nearby atoms or molecular complexes, perturbing their spectra.  We calculate one such effect, involving the ``axion electrodynamics'' relevant to topological insulators, quantitatively, and identify a signature for T violating superconductivity.  We provide an operator framework whereby effects of this kind can be analyzed systematically.
\end{abstract}
\preprint{MIT-CTP/5056}
\maketitle

%%%Introduction%%%

\textit{Introduction}:  Over the past few decades physicists have come to appreciate the importance of increasingly subtle forms of symmetry breaking in materials, often connected with topology and entanglement \cite{haldane, wen, spinIce}.  Many new states of matter characterized by such ``hidden'' symmetry breaking have been proposed theoretically, but concrete, unambiguous experimental manifestations have been relatively sparse.    Many of the proposed states violate some combination of the discrete symmetries $P, T$ \cite{xlqi}.  This opens up the possibility of unusual polarizabilities, generalizing the familiar dielectric and para- or diamagnetic response parameters $\epsilon, \mu$.   Those polarizabilities can support novel electromagnetic effects, which reflect the discrete symmetry breaking directly \cite{spaldin,essin}.   The effects involve virtual two-photon exchange in loops, and are intrinsically quantum-mechanical.  These effects lead to long-range (generalized) Casimir-type forces, also involving spin \cite{milton}, but our estimates make it plausible that they are more easily accessed through spectroscopy.  Two particularly interesting cases, on which we will focus especially, are boundary Chern-Simons models \cite{coh} and chiral superconductors \cite{Mackenzie}. Both these phenomena have attracted much theoretical attention, and experimental signatures of the postulated symmetry breaking should be helpful in validating candidates.   We will also discuss the possibility of searching for fundamental electric dipole moments and provide a systematic operator framework for analyzing other cases of symmetry breaking.

{\it Atmosphere from Axion Electrodynamics\,}: Consider a material whose interaction with the electromagnetic field contains an action term
\begin{equation}\label{basicAction}
\int \, d^3 x dt \, \chi_M(x) \,  \Delta {\cal L}_{\rm axion} ~=~  \, \int d^3 x dt \, \chi_M(x) \, \kappa \, \vec E \cdot \vec B \, ,
\end{equation}
where $\chi_M(x)$ is the characteristic function of the material.   This sort of interaction, an induced Chern-Simons term, was contemplated in \cite{axionED}, and it is realized in topological insulators \cite{xlqi,essin,topIns,topInsExp}, with $\kappa = j {\alpha}$, where $j$ is an odd integer.   (Note that while this is the most direct extrapolation of the bulk effective theory of topological insulators, there could in principle be additional, non-universal contributions to the surface action.  Note also that the overall global P, T symmetry of topological insulators cannot be applied locally at boundaries.)  Since $\vec E \cdot \vec B$ is a total derivative, it does not affect the bulk equations of motion.   But when the spatial region occupied by the material is bounded, surface terms arise \cite{footnote1}.   Specifically, if the plane $z=0$ forms an upper boundary, we will have a surface action
\begin{eqnarray}
&{}& \int d^3 x dt \, \chi_M(x) \, \kappa \, \vec E \cdot \vec B \, \nonumber \\
&{}&\rightarrow \frac{\kappa}{2}  \int dx\, dy\,  dt\,  \epsilon^{3\alpha \beta \gamma} A_\alpha ( x, y, 0 , t) \partial_\beta A_\gamma (x, y, 0, t) . 
\end{eqnarray}
This gives us a two-photon vertex which violates the discrete symmetries $P, T$ locally, while preserving $PT$.   Quantum fluctuations involving this vertex will produce a sort of $P, T$ violating atmosphere above the material.  (See Figure 1.)   The atmosphere induces new kinds of  ``Casimir'' forces on bodies near the material \cite{casimirForces1,casimirForces2,casimirForces3,chiralCasimir,axialCasimir}.  It also induces new kinds of effective interactions within atoms or molecular centers, which affect their spectra.   Such interactions are especially interesting, because in favorable cases the spectra can be measured quite accurately, thus plausibly rendering small symmetry-violating effects accessible.   

\begin{figure}[!htb]
\centering
\includegraphics[height=2.82cm, width=8.2cm, angle=0]{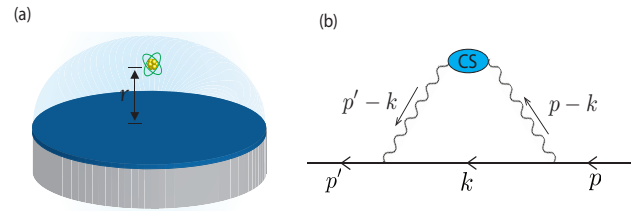}
\caption{(a) Illustration of  quantum atmosphere induced by a Chern-Simons surface. The blue layer corresponds to the top surface described by a Chern-Simons term at $z=0$. Due to quantum fluctuation, time-reversal symmetry breaking effect will be transmitted to the nearby atom at the distance $r$ from the surface. (b) Feynman diagram involving Chern-Simons vertex.  \label{fig:1}}
\end{figure}

Let us analyze the most basic case, that is the interaction of an electron.   By symmetry and dimension counting, the first-order effective $P, T$ violating interaction with an electron, at a distance $r$ from a planar boundary, will take the form
\begin{equation}
{\cal L}_{\rm int.} ~\sim~ \frac{\alpha \kappa}{m r^2} \, \hat n \cdot \vec s \, ,
\end{equation}
where $m, \vec s$ are the electron's mass and spin, and $r, \hat n$ are the distance and normal to the plane.  Expressed using fundamental units only, as in the quoted form for topological insulators, we find the dimensional estimate
\begin{equation}\label{dimensionalEstimate}
{\cal L}_{\rm int.} ~\sim~ \frac{\alpha^2}{mr^2} \,  \hat n \cdot \vec s \,  \approx \,  \, \Bigl(\frac {10 {\rm \, nm.}}{r}\Bigr)^2 \, \frac {e \hat n \cdot s}{m} \, 10 {\rm \ gauss} . 
\end{equation}
Here we have expressed the atmospheric Zeeman-like interaction in a form which allows ready comparison with the Zeeman splitting induced by a magnetic field strength.   Taken at face value, this is comfortably within the estimated sensitivity of magnetometry based on NV centers \cite{magnetometry} -- by many orders of magnitude (but see below).    Note however that we do not generate true magnetic flux, so that SQUID detectors will not register (but see below).   

We can check this estimate by explicit calculation, according to the Feynman digram of Figure 1.   We find \cite{supplematerial}
\begin{equation}\label{bareCalculation}
V(r) ~=~ \frac{\kappa e^2}{128 \pi^2} \, \frac{1}{mr^2} \, \sigma_3 ~\rightarrow \frac{j \alpha^2}{32\pi} \, \frac{1}{mr^2} \, \sigma_3
\end{equation}

One might attempt to generalize this calculation to particles which possess an anomalous magnetic moment (e.g., atomic nuclei), but one encounters an ultraviolet divergence \cite{supplematerial}.   This is not a physical contradiction, because both anomalous magnetic moments and (especially) our assumed action Eqn.\,(\ref{basicAction}) will have form-factors which provide cut-offs.   Also, of course, the virtual photons emitted from the material need not terminate on a single particle.  For these reasons, our estimate Eqn.\,(\ref{dimensionalEstimate}) and the result of our calculation Eqn.\,(\ref{bareCalculation}) should be regarded as encouraging, but applied with care.   Dispersion relations relating spectroscopic splitting to the material's response to photons are included in the supplementary material \cite{supplematerial}.

We can also consider the effect of applying an external electric field.  Importantly, this does not in itself introduce $T$ violation.  If we apply an electric field parallel to the boundary plane, we induce a surface Hall-like current.  A planar current sheet produces a spatially constant (true) magnetic field, which will be aligned (or anti-aligned) with the applied electric field.  To maximize the induced field while avoiding cancellations between contributions from opposite sides of the material, we should use samples with effective surfaces whose linear dimensions are large compared to the distance to the test atom or complex, but small compared to the separation between surfaces.  If we apply an electric field perpendicular to the boundary plane, it induces a surface magnetic charge, and thus again a magnetic field aligned or anti-aligned with the applied electric field, and in the same sense.  The magnitudes of the magnetic fields, for moderate values of the applied electric field, can be quite substantial:
\begin{equation}
B ~\sim~ \kappa E ~\rightarrow~ \alpha E \, \approx \, 10^{-1} \, {\rm gauss} \, \Bigl( \frac{E}{10^4 \frac{\rm V}{\rm cm.}} \Bigr)
\end{equation}
where the progression from general to particular is as previously.  These induced currents and fields were anticipated in \cite{axionED}; here we are adding some context on their connection with symmetry and their possible experimental accessibility.  They are a much more conservative application of the effective theory.

{\it Atmosphere of Superconductors}: The classic signature for superconductivity is the Meissner effect, i.e. exclusion of an applied magnetic field.  This signature is not ideal for discovery work, since the superconducting regions can be small and the superconductivity itself disrupted by magnetism.   Spectroscopic shifts induced by Meissner response to virtual photons can offer an alternative.   Such shifts were calculated in \cite{haakh,skag}, under the assumption of  $T$ symmetry.  Violation of $T$ symmetry can induce splitting between states that are otherwise degenerate. Chiral superconductors are typical examples where time-reversal symmetry is broken due to the finite orbital angular momentum of Cooper pairs \cite{thughs,etaylor}. This leads to a state-dependent magnetic energy shift \cite{supplematerial}
\begin{eqnarray}
\delta \epsilon_n=\sum_m &&\int_0^\infty \frac{d\omega}{2\pi}~\frac{2\epsilon_{mn}}{\epsilon_{mn}^2-\omega^2}\times\nonumber\\
&&{\rm Im}\left\{\langle n| D_1|m\rangle\langle m|  D_2|n\rangle\mathcal H_{12}(z,z;\omega)\right. \nonumber\\
&&\quad\quad\left. +\langle n| D_2|m\rangle\langle m|  D_1|n\rangle\mathcal H_{21}(z,z;\omega)\right\},
\end{eqnarray}
%\begin{eqnarray}
%&&\delta \epsilon_n=\sum_m \langle n| D_1|m\rangle\langle m|  D_2|n\rangle \int_0^\infty d\xi ~\frac{2\epsilon_{mn}\mathcal H_{12}(z,z;i\xi)}{\xi^2+\epsilon_{mn}^2}\nonumber\\
%&&+\sum_m \langle n| D_2|m\rangle\langle m| D_1|n\rangle \int_0^\infty d\xi ~\frac{2\epsilon_{mn}\mathcal H_{21}(z,z;i\xi)}{\xi^2+\epsilon_{mn}^2}
%\end{eqnarray}
where $\vec D$ is the magnetic dipole operator, the coordinates are labelled $1,2,z$, and $\mathcal H$ is the frequency-dependent modification of the magnetic field correlator due to the superconductor.   $T$ violation introduces an imaginary part into $\mathcal H_{12} (= - \mathcal H_{21})$ and leads to an effective interaction which splits states of opposite angular momentum in the $z$ direction \cite{footnote2}.  It mimics, in other words, the effect of a Zeeman interaction with an emergent magnetic field.

{\it Fundamental Electric Dipole Moments}: Apart from spontaneous $P, T$ symmetry breaking in materials, we may also have intrinsic violation.  That possibility is of great interest for fundamental physics \cite{pospelov}.   A generic signature of such violation is the existence of particles having both elementary magnetic dipole moments and (small) elementary electric dipole moments.  (Let us emphasize that this represents physics beyond the ``standard model''.)   A material containing a density $\rho$ of such particles will, in the presence of an applied electric field at temperature $T$, contain a density $\rho g_e \vec E/T$ of aligned spins, and hence an energy density $\bigl( g_m g_e / T \bigr) \rho \vec E \cdot \vec B$.   Thus, we identify an alternative source of our action Eqn.\,(\ref{basicAction}), with $\kappa = \rho g_m g_e / T$.   In this model, it is transparently clear why a normal electric field, by inducing a magnetic dipole density, yields a surface magnetic charge density.   Some possible experimental arrangements to probe intrinsic symmetry breaking effects of this kind were discussed in \cite{dipoleMomentExperiment} from a very different point of view.  Numerically, we have 
\begin{eqnarray}
B &\sim& \rho g_m g_e E/T \, \nonumber \\
&\sim& \Bigl( \frac {\rho} { \frac {10^{22}} {{\rm cm}^3} } \Bigr) \, \frac{g_e}{10^{-26} {\rm e}\,{\rm cm}} \frac{E}{10^6 \frac{\rm V}{\rm cm} }  \frac{10^{-3} {\rm K}}{T} 10^{-12} {\rm gauss}
\end{eqnarray}
where we have inserted the electron gyromagnetic moment, aggressive reference values of the parameters, and a reference value of the electric dipole moment comparable to current limits.  The resulting magnetic field is well within advertised sensitivities \cite{magnetometry}. Note that in this estimate we have assumed a thermal population of the spins, for which the asymmetry is suppressed, due to the tininess of the electric moment energy splitting.

{\it Operator Analysis of Polarizabilities\,}: In constructing effective theories of electromagnetism in condensed matter, there are few principles we can apply {\it a priori}.   Nevertheless, when plausible assumptions and approximations give us tractable theories which contain few parameters, those theories can be very useful in organizing data and planning experiments.   For our purposes, it is instructive to recall that textbooks of electromagnetism commonly introduce just two material-dependent parameters, $\epsilon$ and $\mu$, to describe a wide range of observed behaviors.  They can be considered as coefficients in the Maxwell action
\begin{eqnarray}\label{maxwellParameters}
&{}&\int \, d^3 x dt \, \chi_M(x) \,  \Delta {\cal L}_{\rm Maxwell} \nonumber \\
~&=&~  \, \int d^3 x dt \, \chi_M(x) \, \bigl( \,  \frac{\epsilon}{2} \vec E^2 \, - \, \frac{1}{2 \mu} \vec B^2 \, \bigr) .
\end{eqnarray}
These are the possible terms which satisfy four sorts of conditions:
\begin{enumerate}
\item They are local in space and time, containing only products of fields at the same space-time point.
\item They are invariant under many symmetries: time and space translation, rotation, gauge.
\item They are quadratic in fields and of lowest possible order (i.e., zero) in space and time gradients.
\item They are invariant under $P$ and $T$ symmetry.
\end{enumerate}

Eqn.\,(\ref{basicAction}) is an additional term we can bring in if we drop the last of those conditions.  Aside from symmetry, is also commonly ignored because it does not contribute to the bulk equations of motion, but as we have seen that reason is superficial.  

The third condition is practical rather than fundamental.   Indeed, terms containing higher powers of fields are the meat and potatoes of nonlinear optics \cite{bloembergen}.   But in many circumstances it is appropriate to ignore nonlinear effects.  Also, it is often appropriate to consider external and effective fields which vary smoothly in space in time.   With those ideas in mind, we can get a nice inventory of the possible terms which are quadratic in fields and of lowest order in space and time gradients while consistent with 1.-3. and displaying different $P$, $T$ characters.   We arrive at the following candidate Lagrangian densities:
\begin{itemize}
\item {\it P even, T even}: Maxwell terms, Eqn.\,(\ref{maxwellParameters})
\begin{eqnarray}
{\cal O}_E ~&=&~ {\vec E}^2 \nonumber \\
{\cal O}_B ~&=&~ {\vec B}^2 
\end{eqnarray}
\item {\it P odd, T odd}:  axion electrodynamics, Eqn.\,(\ref{basicAction})
\begin{equation}
{\cal O}_a ~=~ \vec E \cdot \vec B
\end{equation}
\item {\it P even, T odd}: 
\begin{eqnarray}
{\cal O}_1 ~&=&~ \frac{\partial{\vec E}}{\partial t} \cdot \vec E = \frac{\partial}{\partial t}  \frac{1}{2} \vec E^2 \nonumber \\
{\cal O}_2 ~&=&~ \frac{\partial{\vec B}}{\partial t} \cdot \vec B = \frac{\partial}{\partial t}  \frac{1}{2} \vec B^2 \nonumber  \\
{\cal O}_3 ~&=&~ \bigl[ (\nabla \times \vec E) \cdot \vec B \bigr] \nonumber \\
{\cal O}_4 ~&=&~ (\nabla \times \vec B) \cdot \vec E = {\cal O}_3 - \nabla \cdot (\vec E \times \vec B) 
\end{eqnarray}
\item {\it P odd, T even}:  
\begin{eqnarray}
{\cal O}_5 ~&=& \bigl[ (\nabla \times \vec E) \cdot \vec E \bigr] \nonumber \\
{\cal O}_6 ~&=&~ (\nabla \times \vec B) \cdot \vec B \nonumber \\
{\cal O}_7 ~&=&~ \frac{\partial{\vec E}}{\partial t} \cdot \vec B \nonumber \\
{\cal O}_8 ~&=&~ \frac{\partial{\vec B}}{\partial t} \cdot \vec E = \frac{\partial}{\partial t} ( \vec B \cdot \vec E ) - {\cal O}_7
\end{eqnarray}
\end{itemize}
The bracketed terms are redundant, since the Faraday relation $\nabla \times \vec E = - \frac{\partial B}{\partial t}$ holds identically, when one expresses the fields in terms of potentials.   Terms which are total time derivatives do not contribute to the equations of motion or to surface times, while terms which are total space divergences give boundary actions.   Thus in the $P$ even, $T$ odd case we find only a boundary action, corresponding to ${\cal O}_4$, while in the $P$ odd, $T$ even case we get two terms, corresponding to ${\cal O}_6$ and ${\cal O}_7 - {\cal O}_8$, which affect bulk behavior.   These considerations can guide the design of experiments.   For example, to search for a $P$ violating but $T$ invariant atmosphere (and thus, to probe for states of matter with those symmetries) we might first exclude an emergent $\hat n \cdot \vec s$ interaction in a planar geometry, and then look for an emergent $ \hat n_1 \cdot (\hat n_2 \times \vec s)$ interaction in a more complex geometry, involving two characteristic directions.   Upon applying a time-dependent electric field, we may look for an atmospheric magnetic field whose direction changes according to whether the magnitude of $\vec E$ is increasing or decreasing.   That behavior derives from ${\cal O}_7$.  ${\cal O}_5$ and ${\cal O}_6$, which were considered formally in \cite{lipkin}, where they were referred to as ``zilch'', without proposed application. 

% It is interesting to note that, in view of the displayed equations, all but one of these terms can be expressed as a total derivative.  Thus, like ${\cal O}_a$, they do not contribute to the bulk equation of motion.  (The exception, ${\cal O}_6$, is employed in the description of gyrotropic materials.). Nevertheless, spatial divergences will contribute surface actions and quantum atmospheres, and affect the response to external fields, along the lines we discussed previously.    Thus we arrive at a unique relevant $P$ even, $T$ odd term ${\cal O}_4$, and two relevant $P$ odd, $T$ even terms, ${\cal O}_6$ and ${\cal O}_7$.

Note that if we work directly at the level of polarizabilities, rather than actions, we can define contributions corresponding to all eight cases, and also two independent ``axion'' terms.  Thus for example we might write
\begin{eqnarray}
\vec D ~&=&~ c_e \vec E + c_{a1} \vec B + c_1 \frac{\partial \vec E}{\partial t} + c_4 \nabla \times \vec B +  c_5  \nabla \times \vec E + c_8  \frac{\partial \vec B}{\partial t}  \nonumber \\
\vec H ~&=&~ c_b \vec B + c_{a2} \vec E + c_2  \frac{\partial \vec B}{\partial t} + c_3 \nabla \times \vec E + c_6  \nabla \times \vec B + c_7 \frac{\partial \vec E}{\partial t}  \nonumber  \, . \\
&{}&
\end{eqnarray}
After applying the Faraday relation, we have ten independent terms, including the two conventional ones.  The more restrictive Lagrangian approach seems more principled, however.  

Materials that contain chiral molecules can violate $P$ while conserving $T$ intrinsically; indeed, many such so-called gyrotropic materials are well known \cite{gyrotropyMatt}. The recently discovered P-violating Weyl semimetals, which display the chiral magnetic effect in transport, provide another example \cite{armitagereview}.  A possibility for more subtle, spontaneous breaking of this class, which still preserves macroscopic rotation and translation symmetry, could be a non-vanishing correlation of the type $\langle \vec j \cdot \vec s \rangle \neq 0$ between microscopic current and and spin densities which are themselves uncorrelated ($\langle \vec j \rangle = \langle \vec s \rangle = 0$).  Similarly, a non-vanishing correlation of the type $\langle \vec j \cdot \vec \pi \rangle \neq 0$ between microscopic current and polarization densities which are themselves uncorrelated exhibits $P$ even, $T$ odd spontaneous breaking; while a  non-vanishing correlation $\langle \vec s \cdot \vec \pi \rangle \neq 0$ is odd under both $P$ and $T$, but even under $PT$, as we have mentioned before implicitly.  

%%%Summary%%%Summary%%%Summary%%%
\textit{Summary}:  We have discussed how quantum fluctuations, in the presence of a material, produce a kind of atmosphere which can affect the spectra of nearby atoms.  The atmosphere can be probed to diagnose properties of the material, and in particular its symmetry.   We have calculated one effect of this kind, by taking the effective theory based on axion electrodynamics at face value, and found a result that is very large compared to expected experimental sensitivities.    The atmosphere can be influenced in a calculable way by external fields.   We displayed an operator framework in which to discuss these issues systematically, and classified the simplest non-trivial possibilities under stated, broad assumptions.   Our assumptions could be relaxed, for instance to allow crystalline asymmetries, at the cost of bringing in more operators.  The operator analysis suggests how to probe symmetry-breaking atmospheres experimentally, and to parameterize their properties systematically.

\textit{Acknowledgement}: This work was supported by the Swedish Research Council under Contract No. 335-2014-7424.  In addition, FW's work is supported by the U.S. Department of Energy under grant Contract  No. DE-SC0012567 and by the European Research Council under grant 742104.

%%%Reference%%%Reference%%%Reference%%e

\hspace{3mm}

%%%Supplementary%%%Supplementary%%%Supplementary%%%


\begin{thebibliography}{}
\bibitem{footnote1}
One can stabilize the Chern-Simons action on the surface of topological insulators by growing a ferromagnetic thin layer on the top of a topological insulator. See Ref. \cite{xlqi} for more details.
\bibitem{footnote2}
Note that this involves a violation of Lorentz reciprocity, a consequence of $T$ symmetry, which implies $\mathcal H_{ij}(\bold r,\bold r^\prime,\omega)=\mathcal H_{ji}(\bold r^\prime,\bold r,\omega)$.
\end{thebibliography}
\end{document}

\bibliographystyle{unsrt}
%%%%%%%%%%%%%%%%%%%%%%%%%%
\title{Supplemental Materials for ``Quantum Atmospherics for Materials Diagnosis"}
\author{Q.-D. Jiang and F. Wilczek}
%\date{\today}
\maketitle
\tableofcontents
\bigskip

\section{I. Evaluating the one-loop Feynman diagram with Chern-Simons interaction}

In this section, we calculate  the effective potential for an electron in the vicinity of a Chern-Simons surface. The key step is to evaluate the two-photon exchange Feynman diagram shown in the figure \ref{figAppend} . 

\subsection{1. Scattering matrix of the two-photon exchange Feynman diagram}

Consider an electron moving at a distance $r$ above a Chern-Simons (CS) surface at $z=0$. The action has the following form
\begin{eqnarray}
S=\int d^4 x\,\,\left\{ \bar\psi \left[\gamma^\mu(p_\mu-eA_\mu)-m\right]\psi-\frac{1}{4}F_{\mu\nu}F^{\mu\nu}\right\}+\int d^4 x \,\epsilon^{\alpha\beta\rho 3} A_{\alpha}\partial_\beta A_{\rho} \,\delta(x_3).
\end{eqnarray}

We separate the whole action into free part and interaction part, i.e., $S=S_0+S_I$, where
\begin{eqnarray}
\label{freeaction}S_0&=&\int d^4 x \,\, \left\{ \bar\psi \left[\gamma^\mu p_\mu-m\right]\psi -\frac{1}{4}F_{\mu\nu}F^{\mu\nu}\right\};\\
\label{intaction}S_I&=&S_I^a+S_I^b=\int d^4 x \,\, \bar\psi \left(-e\gamma^\mu A_\mu \right)\psi +\int d^4 x \, \epsilon^{\alpha\beta\rho 3} A_{\alpha}\partial_\beta A_{\rho} \,\delta(x_3).
\end{eqnarray}
Note that $S_I^a$ and $S_I^b$, respectively, represent electron-photon vertex and CS vertex.

Now, we can consider the generating function
\begin{eqnarray}
Z=\frac{\int D[\bar\psi,\psi] D[A] \,\, e^{iS_0+iS_I}}{\int D[\bar\psi,\psi] D[A]  \,\, e^{iS_0}}=\frac{\int D[\bar\psi,\psi] D[A]  \,\, e^{iS_0}\left[1+i S_I +\frac{1}{2} (iS_I)^2+\frac{1}{3!}(iS_I)^3+...\right]}{\int D[\bar\psi,\psi] D[A] \,\, e^{iS_0}}
\end{eqnarray}

So the lowest order contribution from the CS plate is a two-photon process: (two electron-photon vertices and one CS vertex)
\begin{eqnarray}
Z=\frac{\int D[\bar\psi,\psi] D[A]  \,\, e^{iS_0}\left[\frac{1}{2}(i\,S_I^a)^2(i\,S_I^b)\right]}{\int D[\bar\psi,\psi] D[A]  \,\, e^{iS_0}}.
\end{eqnarray}

\begin{figure}[!htb]
\centering
\includegraphics[height=2.8cm, width=11cm, angle=0]{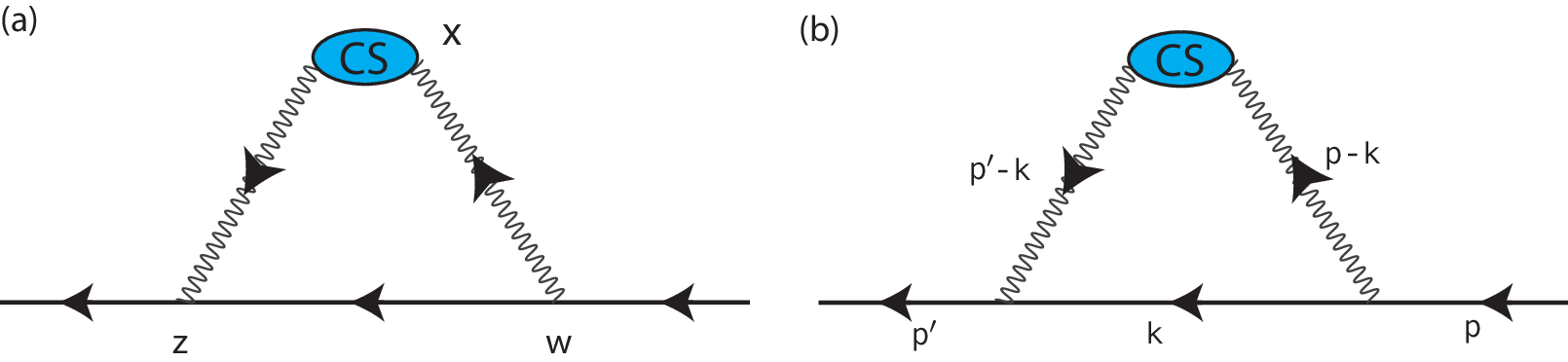}
\caption{The Feynman diagrams in real space (a) and in momentum space (b). \label{figAppend}}
\end{figure}

The relevant Feynman diagram [See the figure \ref{figAppend}.] describe the interaction between the electron and Chern-Simons term can be calculated via 
\begin{eqnarray}\label{scatteringAmplitude}
M=\int d^4 x \int d^4 w \int d^4 z \,\,\bar\psi(z) \,(-ie\gamma^\mu) D_{\mu\alpha}(z-x) \, G(z-w) \,(i\partial_\beta)\, D_{\rho\delta}(x-w)\, \delta(x_3) \epsilon^{\alpha\beta\rho 3}\, (-ie\gamma^{\delta}) \,\psi(w)
\end{eqnarray}
where $G$ and $D$ correspond to Feynman propagators of electron and photon, respectively. 

Substitute the Fourier transformation of the Feynman propagators
\begin{eqnarray}
&&D_{\mu\alpha}(z-x)=\int \frac{d^4 k^{\prime}}{(2\pi)^4} \,\frac{(-i)g_{\mu\alpha}}{{k^{\prime}}^2+i\epsilon} \,e^{-ik^{\prime}(z-x)}\\
&&D_{\rho\delta}(x-w)=\int \frac{d^4 k^{\prime\prime}}{(2\pi)^4} \,\frac{(-i)g_{\rho\delta}}{{k^{\prime\prime}}^2+i\epsilon} \,e^{-ik^{\prime\prime}(x-w)}\\
&&G(z-w)=\int \frac{d^4 k}{(2\pi)^4}\,\frac{i}{\gamma^\mu k_\mu-m+i\epsilon} e^{-ik(z-w)}
\end{eqnarray}
into the above expression Eqn. \eqref{scatteringAmplitude}, and one can obtain:

\begin{eqnarray}\label{amplitudeM}
M&=&\int d^4x \int d^4w \int d^4z\,\, \delta(x_3)\times \bar u(p^{\prime})e^{ip^{\prime} z}(ie\gamma^{\mu})\times\nonumber\\
&&\int \frac{d^4 k^{\prime}}{(2\pi)^4}D_{\mu\alpha}(k^{\prime})e^{-i k^{\prime}(z-x)}\times \,(i\partial_\beta)\, \int \frac{d^4 k^{\prime\prime}}{(2\pi)^4} D_{\rho\delta}(k^{\prime\prime})e^{-i k^{\prime\prime}(x-w)}\epsilon^{\alpha\beta\rho 3}\times \int \frac{d^4 k}{(2\pi)^4}  G(k) e^{-i k(z-w)}\,\, e^{-ipw} (ie\gamma^{\delta})u(p)\nonumber\\
&=&\bar u(p^{\prime}) (ie\gamma^\mu)\,\int dx_0dx_1dx_2dx_3 \int dw_0 dw_1 dw_2 dw_3 \int dz_0 dz_1 dz_2 dz_3\,\, \delta(x_3)\times\nonumber\\
&&\int \frac{d^4 k}{(2\pi)^4} \int \frac{d^4 k^{\prime}}{(2\pi)^4} \int \frac{d^4 k^{\prime\prime}}{(2\pi)^4} \left[D_{\mu\alpha}(k^{\prime}) \times (k_\beta^{\prime\prime})\times D_{\rho\delta}(k^{\prime\prime})\right]\epsilon^{\alpha\beta\rho 3} G(k) (ie\gamma^\delta)e^{i(k^\prime-k^{\prime\prime})x}e^{i(k^{\prime\prime} +k-p)w}e^{-i(k+k^{\prime}-p^{\prime})z} u(p)\nonumber\\
&=&\frac{1}{2\pi}\bar u(p^{\prime}) (ie \gamma^\mu) \int d^4 k \int d^4 k^{\prime} \int d^4 k^{\prime\prime} \left[D_{\mu\alpha}(k^{\prime}) \times (k_\beta^{\prime\prime})\times D_{\rho\delta}(k^{\prime\prime})\right]\epsilon^{\alpha\beta\rho 3} G(k) (ie\gamma^\delta) \times\nonumber\\
&& \delta(k^{\prime}-k^{\prime\prime})_{0,1,2}\,\,\delta(k^{\prime}+k-p^{\prime}) \,\,\delta(k^{\prime\prime}+k-p)u(p)\,\,
\nonumber\\
&=&\frac{1}{2\pi}\,\bar u(p^{\prime}) \,(ie\gamma^\mu) \int d^4 k\, \delta(p^{\prime}-p)_{0,1,2}\left[D_{\mu\alpha}(p^{\prime}-k)\times (p-k)_\beta \times D_{\rho\delta}(p-k)\right]\epsilon^{\alpha\beta\rho 3} G(k) (ie\gamma^\delta) \times \,u(p)
\end{eqnarray}

In the Feynman gauge, photon's propagator is diagonal. So the scattering amplitude is
\begin{eqnarray}
\begin{aligned}
M=&\frac{1}{2\pi}\, \delta(p^{\prime}-p)_{0,1,2}\,\,\bar u(p^{\prime})(ie\gamma^\mu)\int d^4 k \times\\
&\frac{(-i) g_{\mu\mu}}{(p^{\prime}-k)^2}\times\frac{(-i)g_{\rho\rho}}{(p-k)^2}\times (p-k)_\beta \times\frac{i}{\gamma^\nu k_\nu-m}\,\epsilon^{\mu\beta\rho 3}(ie\gamma^\rho)\, u(p)
\end{aligned}
\end{eqnarray}

We can explicitly write down all possible terms according to the value of $\beta$ in the above formula.

(i) $\beta=1$ term:

\begin{eqnarray}
\begin{aligned}
M_1=&-\frac{i}{2\pi}\,\delta(p^{\prime}-p)_{0,1,2}\,\bar u(p^{\prime})\,(e^2\gamma^0)\int d^4 k\times
\frac{1}{(p-k)^2}\times\frac{1}{(p^{\prime}-k)^2}\times \,\frac{(p-k)_1}{\gamma^\nu k_\nu-m} \gamma^2\times \, u(p)\\
&+\frac{i}{2\pi}\,\delta(p^{\prime}-p)_{0,1,2}\,\bar u(p^{\prime})\,(e^2\gamma^2)\int d^4 k\times
\frac{1}{(p-k)^2}\times\frac{1}{(p^{\prime}-k)^2}\times \,\frac{(p-k)_1}{\gamma^\nu k_\nu-m} \gamma^0\times \, u(p)\\
=&-\frac{ie^2}{2\pi}\,\delta(p^{\prime}-p)_{0,1,2}\,\bar u(p^{\prime})\times\\
&\int d^4 k\,
\frac{1}{(p-k)^2}\times\frac{1}{(p^{\prime}-k)^2} \times \,\frac{\gamma^0(p-k)_1\,(\gamma^\nu k_\nu+m)\gamma^2-\gamma^2(p-k)_1\,(\gamma^\nu k_\nu+m)\gamma^0}{k^2-m^2}\times \, u(p)
\end{aligned}
\end{eqnarray}

%Be careful about the red parts
(ii) $\beta=2$ term:

\begin{eqnarray}
\begin{aligned}
M_2=&\frac{ie^2}{2\pi}\,\delta(p-p)_{0,1,2}\,\bar u(p^{\prime})\times\\
& \int d^4 k \,
\frac{1}{(p-k)^2}\times\frac{1}{(p^{\prime}-k)^2}\times \,\frac{\gamma^0(p-k)_2\,(\gamma^\nu k_\nu+m)\gamma^1-\gamma^1(p-k)_2\,(\gamma^\nu k_\nu+m)\gamma^0}{k^2-m^2}\times \, u(p)
\end{aligned}
\end{eqnarray}

(iii) $\beta=0$ term:
\begin{eqnarray}
\begin{aligned}
M_3=&-\frac{i e^2}{2\pi}\,\delta(p^{\prime}-p)_{0,1,2}\,\bar u(p^{\prime})\times\\
&\int d^4 k \,
\frac{1}{(p-k)^2}\times\frac{1}{(p^{\prime}-k)^2}\times \,\frac{\gamma^1(p-k)_0(\gamma^\nu k_\nu+m)\gamma^2-\gamma^2(p-k)_0(\gamma^\nu k_\nu+m)\gamma^1}{k^2-m^2}\times \, u(p)
\end{aligned}
\end{eqnarray}
\\

\subsection{2. Calculation of the integrals in the scattering matrix}

\underline{First of all, let's perform Feynman parametrization to simplify the denominator.}

Using Feynman parametrization trick $\frac{1}{ABC}=2\int_0^1 du_1\int_0^{u_1} du_2 \frac{1}{\left[u_2A+(u_1-u_2)B+(1-u_1)C\right]^3}$, one can obtain
\begin{eqnarray}
\begin{aligned}
\frac{1}{(p-k)^2}\times\frac{1}{(p^{\prime}-k)^2}\times\frac{1}{k^2-m^2}&=2\int_0^1 du_1 \int_0^{u_1} du_2 \frac{1}{\left[u_2(p-k)^2+(u_1-u_2)(p^{\prime}-k)^2+(1-u_1)(k^2-m^2)\right]^3}\\&=2\int_0^1 du_1 \int_0^{u_1} du_2 \frac{1}{D^3}
\end{aligned}
\end{eqnarray}

Here,

\begin{eqnarray}\label{feynmanPara}
\begin{aligned}
D=&u_2(p-k)^2+(u_1-u_2)(p^{\prime}-k)^2+(1-u_1)(k^2-m^2)\\
=& u_2 \left[(p-k)_0^2-(p-k)_1^2-(p-k)_2^2-(p-k)_3^2\right]+(u_1-u_2)\left[(p^{\prime}-k)_0^2-(p^{\prime}-k)_1^2-(p^{\prime}-k)_2^2-(p^{\prime}-k)_3^2\right]\\
&+(1-u_1)\left(k_0^2-k_1^2-k_2^2-k_3^2-m^2\right)\\
=&-u_2(p_3-k_3)^2+u_2(p_3^{\prime}-k_3)^2+u_1\left[(p^{\prime}-k)_0^2-(p^{\prime}-k)_1^2-(p^{\prime}-k)_2^2-(p^{\prime}-k)_3^2\right]\\
&+(1-u_1)\left(k_0^2-k_1^2-k_2^2-k_3^2-m^2\right)\\
=&2u_2(p_3-p_3^{\prime}) k_3+u_1\left[({p_0^{\prime}}^2-2p_0^{\prime}k_0)-({p_1^{\prime}}^2-2p_1^{\prime}k_1)-({p_2^{\prime}}^2-2p_2^{\prime}k_2)-({p_3^{\prime}}^2-2p_3^{\prime}k_3)\right]\\
&+\left(k_0^2-k_1^2-k_2^2-k_3^2\right)-(1-u_1)m^2\\
=&(k_0^2-2u_1p_0^{\prime} k_0)-(k_1^2-2u_1p_1^{\prime} k_1)-(k_2^2-2u_1p_2^{\prime} k_2)
-\left[k_3^2-2u_1p_3^{\prime} k_3+2u_2(p_3^{\prime}-p_3)k_3\right]\\&+
u_1({p_0^{\prime}}^2-{p_1^{\prime}}^2-{p_2^{\prime}}^2-{p_3^{\prime}}^2)-(1-u_1)m^2\\
=&(k_0^2-2u_1p_0^{\prime} k_0)-(k_1^2-2u_1p_1^{\prime} k_1)-(k_2^2-2u_1p_2^{\prime} k_2)
-\left[k_3^2-2u_1p_3^{\prime} k_3+2u_2(p_3^{\prime}-p_3)k_3\right]\\&+
u_1 m^2-(1-u_1)m^2\\
=&(k_0-u_1p_0^{\prime})^2-(k_1-u_1p_1^{\prime})^2-(k_2-u_1p_2^{\prime})^2-\left[k_3-u_1p_3^{\prime}+u_2(p_3^{\prime}-p_3)\right]^2\\
&+(2u_1-1)m^2-
(u_1p_0^{\prime})^2+(u_1p_1^{\prime})^2+(u_1p_2^{\prime})^2+\left[-u_1p_3^{\prime}+u_2(p_3^{\prime}-p_3)\right]^2\\
=&(k_0-u_1p_0^{\prime})^2-(k_1-u_1p_1^{\prime})^2-(k_2-u_1p_2^{\prime})^2-\left[k_3-u_1p_3^{\prime}+u_2(p_3^{\prime}-p_3)\right]^2\\
&+(-u_1^2+2u_1-1)m^2-
u_1^2{p_3^{\prime}}^2+((u_1-u_2)p_3^{\prime}+u_2p_3)^2\\
=&l_0^2-l_1^2-l_2^2-l_3^2-(1-u_1)^2m^2-u_1^2{p_3^{\prime}}^2+\left[-u_1p_3^{\prime}+u_2(p_3^{\prime}-p_3)\right]^2\\
=&l_0^2-l_1^2-l_2^2-l_3^2-T^2
\end{aligned}
\end{eqnarray}
where $T^2=(1-u_1)^2m^2+u_1^2{p_3^{\prime}}^2-\left[u_1p_3^{\prime}-u_2(p_3^{\prime}-p_3)\right]^2$. We have used substitution of variables
$l_0=k_0-u_1 p_0^{\prime}$, $l_1=k_1-u_1 p_1^{\prime}$, $l_2=k_2-u_1 p_2^{\prime}$, $l_3=k_3-u_1 p_3^{\prime}+u_2 (p_3^{\prime}-p_3)$, and on-shell condition of the external legs of electrons in Eqn. \eqref{feynmanPara}. 
\\

\underline{Second, let's make some simplification of the numerator.}

The numerator in $M_1$ is

\begin{eqnarray}
\begin{aligned}
&\gamma^0(p-k)_1\,(\gamma^\nu k_\nu+m)\gamma^2-\gamma^2(p-k)_1\,(\gamma^\nu k_\nu+m)\gamma^0\\=&
\gamma^0\gamma^2(p_1-k_1)\left[-\gamma^2(\slashed{k}+m)\gamma^2+\gamma^0 (\slashed{k}+m)\gamma^0\right]\\=&
\gamma^0\gamma^1\gamma^2(\slashed{p}_1-\slashed{k}_1)\left[-\gamma^2(\slashed{k}+m)\gamma^2+\gamma^0 (\slashed{k}+m)\gamma^0\right]\\=&
2\gamma^0\gamma^1\gamma^2(\slashed{p}_1-\slashed{k}_1)\left[m-\slashed{k}_1-\slashed{k}_3\right]
\end{aligned}
\end{eqnarray}

The numerator in $M_2$ is

\begin{eqnarray}
\begin{aligned}
&-\left[\gamma^0(p-k)_2\,(\gamma^\nu k_\nu+m)\gamma^1-\gamma^1(p-k)_2\,(\gamma^\nu k_\nu+m)\gamma^0\right]\\=&
-\gamma^0\gamma^1\gamma^2(\slashed{p}_2-\slashed{k}_2)\left[\gamma^1(\slashed{k}+m)\gamma^1-\gamma^0 (\slashed{k}+m)\gamma^0\right]\\=&
2\gamma^0\gamma^1\gamma^2(\slashed{p}_2-\slashed{k}_2)\left[m-\slashed{k}_2-\slashed{k}_3\right]
\end{aligned}
\end{eqnarray}

The numerator in $M_3$ is

\begin{eqnarray}
\begin{aligned}
&\gamma^1(p-k)_0(\gamma^\nu k_\nu+m)\gamma^2-\gamma^2(p-k)_0(\gamma^\nu k_\nu+m)\gamma^1\\=&
-\gamma^0\gamma^1\gamma^2(\slashed{p}_0-\slashed{k}_0)\left[\gamma^2(\slashed{k}+m)\gamma^2+\gamma^1 (\slashed{k}+m)\gamma^1\right]\\=&
2\gamma^0\gamma^1\gamma^2(\slashed{p}_0-\slashed{k}_0)\left[m-\slashed{k}_0-\slashed{k}_3\right]
\end{aligned}
\end{eqnarray}

If we add up $M_1$, $M_2$, $M_3$, the total numerator is 
\begin{eqnarray}
\begin{aligned}
&2m\gamma^0\gamma^1\gamma^2\left[(\slashed{p}_0-\slashed{k}_0)+(\slashed{p}_1-\slashed{k}_1)+(\slashed{p}_2-\slashed{k}_2)\right]\\&-
2\gamma^0\gamma^1\gamma^2 \left[(\slashed{p}_0-\slashed{k}_0)(\slashed{k}_0+\slashed{k}_3)+ (\slashed{p}_1-\slashed{k}_1)(\slashed{k}_1+\slashed{k}_3)+(\slashed{p}_2-\slashed{k}_2)(\slashed{k}_2+\slashed{k}_3)\right]
\end{aligned}
\end{eqnarray}

\underline{Next, we regroup the numerator into four parts.}

a.
\begin{eqnarray}
t_1=2 m\gamma^0 \gamma^1\gamma^2 \left[(\slashed{p}_0-\slashed{k}_0)+(\slashed{p}_1-\slashed{k}_1)+(\slashed{p}_2-\slashed{k}_2)\right]
\end{eqnarray}

b.
\begin{eqnarray}
t_2=2 \gamma^0 \gamma^1\gamma^2 (k_0^2-k_1^2-k_2^2)
\end{eqnarray}

c.
\begin{eqnarray}
t_3=-2 \gamma^0 \gamma^1\gamma^2 \left[(\slashed{p}_0+\slashed{p}_1+\slashed{p}_2)\slashed{k}_3-(\slashed{k}_0+\slashed{k}_1+\slashed{k}_2)\slashed{k}_3\right]
\end{eqnarray}

d.
\begin{eqnarray}
t_4=-2 \gamma^0 \gamma^1\gamma^2 \left[p_0k_0-p_1k_1-p_2k_2\right]
\end{eqnarray}

The integral that we need to calculate becomes
\begin{eqnarray}
M=-\frac{ie^2}{2\pi}\delta(p^{\prime}-p)_{0,1,2}\times 2 \int_0^1 du_1\int_0^{u_1} du_2 \int d^4l \frac{t_1+t_2+t_3+t_4}{\left[l_0^2-l_1^2-l_2^2-l_3^3-T^2\right]^3},
\end{eqnarray}
where $T^2=\alpha^2-\left[u_1 p_3^{\prime}-u_2(p_3^{\prime}-p_3)\right]^2$ with $\alpha^2=(1-u_1)^2 m^2+u_1^2 {p_3^{\prime}}^2$.

In the following, we will often use the typical integral of momentum:
\begin{eqnarray}
\begin{aligned}
&\int d^4 l\frac{1}{(l_0^2-l_1^2-l_2^2-l_3^2-T^2+i\epsilon)^3} \,\text{(Wick rotation:  $l_0 \rightarrow il_0$)}\\&=
-i\int d^4l \frac{1}{(l_0^2+l_1^2+l_2^2+l_3^2+T^2)^3}\\&=
-i2\pi^2\int_0^{\infty} d l \frac{l^3}{(l^2+T^2)^3}=-i\frac{\pi^2}{2}\frac{1}{T^2}
\end{aligned}
\end{eqnarray}
\\

We will often use variable substitution
$l_0=k_0-u_1p_0^{\prime}$, $l_1=k_1-u_1p_1^{\prime}$, $l_2=k_2-u_1p_2^{\prime}$, $l_3=k_3-u_1p_3^{\prime}+u_2(p_3^{\prime}-p_3)$;
and then perform Wick rotation $l_0\mapsto il_0$ in the following context.  In addition, we use the relations $p_0^{\prime}=p_0$, $p_1^{\prime}=p_1$, $p_2^{\prime}=p_2$ due to the $\delta(p^{\prime}-p)_{0,1,2}$ function in $M$. Now, we can calculate the following terms based on four different types of numerators. 

a.
\begin{eqnarray}
\begin{aligned}
t_1=&2 m\gamma^0 \gamma^1\gamma^2 \left[(\slashed{p}_0^{\prime}-\slashed{k}_0)+(\slashed{p}_1^{\prime}-\slashed{k}_1)+(\slashed{p}_2^{\prime}-\slashed{k}_2)\right]\\&=
-2m\gamma^0 \gamma^1\gamma^2 \left[(\slashed{l}_0+u_1 \slashed{p}_0^{\prime}-\slashed{p}_0^{\prime})+(\slashed{l}_1+u_1 \slashed{p}_1^{\prime}-\slashed{p}_1^{\prime})+ (\slashed{l}_2+u_1 \slashed{p}_2^{\prime}-\slashed{p}_2^{\prime})\right]\\&=
2 m\gamma^0 \gamma^1\gamma^2 (1-u_1)\left[\slashed{p}_0^{\prime}+\slashed{p}_1^{\prime}+\slashed{p}_2^{\prime}\right] \text{( dropped odd power of $l_\mu$)}\\&=
2m(1-u_1)\left[\slashed{p}_0^{\prime}+\slashed{p}_1^{\prime}+\slashed{p}_2^{\prime}\right]\gamma^0\gamma^1\gamma^2\,\,\text{(removed $\slashed{p}_\mu^{\prime}$ to the front)}\\&=
2m(1-u_1)\left[m-\slashed{p}_3^{\prime}\right]\gamma^0\gamma^1\gamma^2\,\,\text{(used on-shell condition of Dirac equation $\bar u(p^{\prime})(\slashed{p}^{\prime}-m)=0$)}\\&
\approx 2m^2 (1-u_1) \gamma^0 \gamma^1\gamma^2 \,\,\text{(in nonrelativistic limit $ m>>p_1,p_2,p_3$)}
\end{aligned}
\end{eqnarray}

b.
\begin{eqnarray}
\begin{aligned}
t_2=&2\gamma^0\gamma^1\gamma^2\left[(l_0+u_1p_0^{\prime})^2-(l_1+u_1 p_1^{\prime})^2-(l_2+u_1p_2^{\prime})^2\right]\\&=
2\gamma^0\gamma^1\gamma^2\left[l_0^2-l_1^2-l_2^2+u_1^2({p_0^{\prime}}^2 - {p_1^{\prime}}^2- {p_2^{\prime}}^2)\right]\text{( dropped odd power of $l_\mu$, and Wick rotation $\rightarrow$)}\\&=
2\gamma^0\gamma^1\gamma^2\left[-l_0^2-l_1^2-l_2^2+u_1^2({p_0^{\prime}}^2 - {p_1^{\prime}}^2- {p_2^{\prime}}^2)\right]\\&=
-2\gamma^0\gamma^1\gamma^2(l_0^2+l_1^2+l_2^2)+2\gamma^0\gamma^1\gamma^2u_1^2\left({p_0^{\prime}}^2 - {p_1^{\prime}}^2- {p_2^{\prime}}^2\right)\\&=
-2\gamma^0\gamma^1\gamma^2(l_0^2+l_1^2+l_2^2)+2\gamma^0\gamma^1\gamma^2u_1^2\left(m^2+{p_3^{\prime}}^2\right)
\end{aligned}
\end{eqnarray}

The first term $2\gamma^0\gamma^1\gamma^2(-l_0^2+l_1^2+l_2^2)$ contributes to the total scattering amplitude as
\begin{eqnarray}
\begin{aligned}
&-\frac{ie^2}{2\pi}\delta(p^{\prime}-p)_{0,1,2}\times 2\int_0^1 du_1 \int_0^{u_1} du_2 (-i)\int d^4 l \frac{-2\gamma^0\gamma^1\gamma^2(l_0^2+l_1^2+l_2^2)}{\left(l_0^2+l_1^2+l_2^2+l_3^2+T^2\right)^3}\\&=
\frac{e^2}{\pi}\delta(p^{\prime}-p)_{0,1,2}\times 2\gamma^0\gamma^1\gamma^2\int_0^1 du_1 \int_0^{u_1} du_2  \times\frac{3}{4} \int_0^{\infty} dl\, (2\pi^2) \frac{l^5}{\left(l^2+T^2\right)^3}\\&=
e^2\pi \delta(p^{\prime}-p)_{0,1,2}\times\gamma^0\gamma^1\gamma^2 \int_0^1 du_1 \int_0^{u_1} du_2 \times \frac{3}{2}\Gamma(0)\\&=
\frac{3\pi e^2}{4}\delta(p^{\prime}-p)_{0,1,2}\gamma^0\gamma^1\gamma^2\Gamma(0)
\end{aligned}
\end{eqnarray}
This term is independent of scattering momentum, thus does not contribute to the effective potential. 

Now, we can consider the second term
\begin{eqnarray}
\begin{aligned}
&2\gamma^0\gamma^1\gamma^2u_1^2\left(m^2+{p_3^{\prime}}^2\right) \,\,\text{(in non-relativistic limit)}\\&\approx
2\gamma^0\gamma^1\gamma^2u_1^2 m^2
\end{aligned}
\end{eqnarray}

c.
\begin{eqnarray}
\begin{aligned}
t_3&=2\gamma^0\gamma^1\gamma^2 \left[(\slashed{k}_0-\slashed{p}_0^{\prime})+(\slashed{k}_1-\slashed{p}_1^{\prime})+(\slashed{k}_2-\slashed{p}_2^{\prime})\right]\slashed{k}_3\\&=
2\gamma^0\gamma^1\gamma^2 \left[(\slashed{l}_0+u_1 \slashed{p}_0^{\prime}- \slashed{p}_0^{\prime})+(\slashed{l}_1+u_1 \slashed{p}_1^{\prime}- \slashed{p}_1^{\prime})+(\slashed{l}_2+u_1 \slashed{p}_2^{\prime}- \slashed{p}_2^{\prime})\right]\times\left[\slashed{l}_3+u_1 \slashed{p}_3^{\prime}-u_2( \slashed{p}_3^{\prime}-\slashed{p}_3)\right]\\&=
-2\gamma^0\gamma^1\gamma^2 (1-u_1)u_1\left(\slashed{p}_0^{\prime}+\slashed{p}_1^{\prime}+\slashed{p}_2^{\prime}\right)\slashed{p}_3^{\prime}+2\gamma^0\gamma^1\gamma^2(1-u_1)u_2 \left(\slashed{p}_0^{\prime}+\slashed{p}_1^{\prime}+\slashed{p}_2^{\prime}\right)(\slashed{p}_3^{\prime}-\slashed{p}_3)\\&=
-2(1-u_1)u_1\left[\gamma^1\gamma^2\gamma^3 p_0^{\prime} p_3^{\prime}+\gamma^0\gamma^2\gamma^3p_1^{\prime}p_3^{\prime}-\gamma^0\gamma^1\gamma^3p_2^{\prime}p_3^{\prime}\right]\\&\,\,\,+
2(1-u_1)u_2\left[\gamma^1\gamma^2\gamma^3 p_0^{\prime} (p_3^{\prime}-p_3) +\gamma^0\gamma^2\gamma^3p_1^{\prime}(p_3^{\prime}-p_3)-\gamma^0\gamma^1\gamma^3p_2^{\prime}(p_3^{\prime}-p_3)\right]\\&=
-2(1-u_1)u_1\left[\gamma^0\gamma^2\gamma^3p_1^{\prime}p_3^{\prime}-\gamma^0\gamma^1\gamma^3p_2^{\prime}p_3^{\prime}\right]+
2(1-u_1)u_2\left[\gamma^0\gamma^2\gamma^3p_1^{\prime}(p_3^{\prime}-p_3)-\gamma^0\gamma^1\gamma^3p_2^{\prime}(p_3^{\prime}-p_3)\right]\\&=
-2(1-u_1)u_1 A+2(1-u_1)u_2 B
\end{aligned}
\end{eqnarray}

If we want to calculate $\int_0^1 du_1 \int_0^{u_1} du_2\,\frac{t_3}{T^2}$, we need to perform the following two integrals:

\begin{eqnarray}
&&\int_0^{u_1} du_2 \frac{u_2}{\alpha^2-\left[u_1p_3^{\prime}-u_2(p_3^{\prime}-p_3)\right]^2} \,\,\left[2(1-u_1)B\right]\\
&&\int_0^{u_1} du_2 \frac{1}{\alpha^2-\left[u_1p_3^{\prime}-u_2(p_3^{\prime}-p_3)\right]^2} \,\,\left[-2(1-u_1)u_1 A\right]
\end{eqnarray}
where $\alpha^2=(1-u_1)^2 m^2+u_1^2{p_3^{\prime}}^2$. 

Because $\int_0^{u_1} du_2 \frac{u_2}{\alpha^2-\left[u_1p_3^{\prime}-u_2(p_3^{\prime}-p_3)\right]^2}=\frac{u_1^2}{\alpha^2}\cdot\frac{p_3^{\prime}}{p_3^{\prime}-p_3}$, and $\int_0^{u_1} du_2 \frac{1}{\alpha^2-\left[u_1p_3^{\prime}-u_2(p_3^{\prime}-p_3)\right]^2}=\frac{u_1}{\alpha^2}$. Then, you will find these two integrals exactly canceled with each other.  So we don't need to consider the term c anymore. Note that these integrals are calculate under the assumption  $p_1\,,p_2\,,p_3<<m$. 

There is another way to prove the vanish of $t_3$ term by invoking the symmetry of the integral. With the substitution $u_2\rightarrow u_1-u_2$, the whole integral remains unchanged. So one can use the combination $\frac{1}{2}\left(u_2+u_1-u_2\right)=\frac{u_1}{2}$ to represent $u_2$. Remember $p_3^{\prime}=-p_3$, then one can show the two terms in $t_3$ cancel out with each other.

d.
\begin{eqnarray}
\begin{aligned}
t_4=&-2\gamma^0\gamma^1\gamma^2\left[p_0^{\prime}k_0-p_1^{\prime}k_1-p_2^{\prime}k_2\right]\\&=
-2\gamma^0\gamma^1\gamma^2\left[p_0^{\prime}(l_0+u_1 p_0^{\prime})-p_1^{\prime}(l_1+u_1 p_1^{\prime})-p_2^{\prime}(l_2+u_1 p_2^{\prime})\right]\\&=
-2\gamma^0\gamma^1\gamma^2u_1\left({p_0^{\prime}}^2-{p_1^{\prime}}^2-{p_2^{\prime}}^2\right)\\&=
-2\gamma^0\gamma^1\gamma^2u_1\left(m^2+{p_3^{\prime}}^2\right)
\\&
\approx -2\gamma^0\gamma^1\gamma^2u_1 m^2
\end{aligned}
\end{eqnarray}

Add up $t_1$, $t_2$ and $t_4$, and the numerator becomes
\begin{eqnarray}
t_1+t_2+t_4=2m^2(1-u_1+u_1^2-u_1)\gamma^0\gamma^1\gamma^2 =2m^2(1-u_1)^2\gamma^0\gamma^1\gamma^2
\end{eqnarray}

Consider $t_1+t_2+t_4$, we need to calculate the integral:
\begin{eqnarray}
\begin{aligned}
&\left[2m^2\gamma^0\gamma^1\gamma^2\right] \int_0^1 d u_1 \int_0^{u_1} du_2 (-\frac{i\pi^2}{2})\frac{(1-u_1)^2}{T^2} \\&=
\left[2m^2\gamma^0\gamma^1\gamma^2\right] \left(-\frac{i\pi^2}{2}\right) \int_0^1 d u_1 \frac{u_1(1-u_1)^2}{\alpha^2} \\&=
\left[(-i\pi^2)\gamma^0\gamma^1\gamma^2  \right]\frac{m^4-\pi m^3p_3^{\prime}-4m^2{p_3^{\prime}}^2 +3m^2{p_3^{\prime}}^2\, log(\frac{m}{p_3^{\prime}})^2}{2m^4}
\end{aligned}
\end{eqnarray}

We collect all the gradients and only care about the off-diagonal scattering amplitude, which is
\begin{eqnarray}
\begin{aligned}
M&=\delta(p^{\prime}-p)_{0,1,2}\,\bar u\,\left(-\frac{ie^2}{2\pi}\right)\times\left[(-i\pi^2)\gamma^0\gamma^1\gamma^2  \right] \,\left(-\frac{\pi}{2}\right)\frac{p_3^{\prime}}{m}\times u\\&=\delta(p^{\prime}-p)_{0,1,2}\, \bar u\,\left(\frac{\pi^2 e^2}{4}\right)\gamma^0\gamma^1\gamma^2 \frac{p_3^{\prime}}{m}\times u
\end{aligned}
\end{eqnarray}

In non-relativistic limit, $u\rightarrow \left(\xi,\,\, \frac{\bold p\cdot \sigma}{2m} \xi\right)^T$, to the first order $p_3^{\prime}/m$,  the scattering amplitude for spin up/down electron is
\begin{eqnarray}
M=\delta(p^{\prime}-p)_{0,1,2}\,\,i \,\xi^{\dagger} \left(-\frac{\pi^2 e^2}{4}\right) \frac{p_3^{\prime}}{m} \sigma_3\, \xi
\end{eqnarray}

In the scattering process, the transferred momentum is $\tilde{p}=(0,0,0,2p_3^{\prime})$. Fourier transform the scattering matrix, we can get the effective interaction
\begin{eqnarray}
\begin{aligned}
V(r)&=\frac{1}{(2\pi)^4}\int_{-\infty}^{\infty}\int_{-\infty}^{\infty} \int_{-\infty}^{\infty} d\tilde{p}_0\,d\tilde{p}_1\, d\tilde{p}_2\,\, \delta(\tilde{p})_{0,1,2}\,\,e^{-i\tilde{p}_0x_0+i\tilde{p}_1 x_1+i\tilde{p}_2 x_2}\,\,\times2\,\int_0^{\infty} d\tilde{p}_3\,\,\left(-\frac{\pi^2 e^2}{4}\right) \frac{p_3^{\prime}}{m} \sigma_3 e^{i\tilde{p}_3 r}\\&=
\frac{1}{16\pi^4}\,\left(-\frac{\pi^2 e^2}{4}\right) \times 4\int_0^{\infty} dp_3^{\prime}\,\frac{p_3^{\prime}}{m}\,\sigma_3\, e^{2ip_3^{\prime}r}\\&=
-\frac{e^2}{16\pi^2}\int_0^{\infty} dp_3^{\prime}\,\frac{p_3^{\prime}}{m}\,\sigma_3\, e^{2ip_3^{\prime}r}\,\,\text{(Wick rotation $p_3^{\prime}\rightarrow ip_3^{\prime}$)}\\&=
\frac{e^2}{16\pi^2}\int_0^{\infty} dp_3^{\prime}\,\frac{p_3^{\prime}}{m}\,\sigma_3\, e^{-2p_3^{\prime}r}\\&=
\frac{e^2}{64\pi^2}\frac{1}{mr^2}\sigma_{3}
\end{aligned}
\end{eqnarray}

Note that, in our calculation, the coefficient before Chern-Simons term is 1. For the surface of a topological insulator, we can put the coefficient as $\frac{\kappa}{2}=\frac{j\alpha}{2}$, where $j$ is an odd number. Thus, our final result becomes 
\begin{eqnarray}
\begin{aligned}
V(r)&=j\frac{\kappa}{2}\times \frac{e^2}{64\pi^2}\frac{1}{mr^2}\sigma_{3}\\&=
j\frac{\alpha^2}{32\pi}\frac{1}{mr^2}\sigma_3
\end{aligned}
\end{eqnarray}
\\

\subsection{3. Considering the contribution from the anomalous magnetic moment}
When one attempts to generalize the above calculation to particles with an anomalous magnetic moment (e.g. atomic nuclei), one encounters ultraviolet divergence.  The ultraviolet divergence can be seen by calculating two-photon interaction terms involving anomalous magnetic moment.
In order to take anomalous magnetic moment into account, one needs add an additional term $S_I^c$ in the interaction part of the original action, i.e.,  $S_I$ in Eq. \eqref{intaction}: 
\begin{eqnarray}
S_I^c=\int d^4 x~g_A~\bar \psi\,\sigma^{\mu\nu}F_{\mu\nu}\,\psi\quad,
\end{eqnarray}
where the Pauli matrices $\sigma^{\mu\nu}\equiv \frac{i}{2}[\gamma^\mu,\,\gamma^\nu]$ and $g_A$ characterizes the magnitude of anomalous magnetic moment.

There are two types of two-photon scattering amplitudes containing anomalous magnetic moment term:
\\
The first type scattering amplitude includes one CS vertex and two anomalous magnetic moment vertices $S_I^bS_I^cS_I^c$, which can be calculated from the following integral:
\begin{eqnarray}\label{A1}
M_{A1}&=&\int d^4x \int d^4 w\int d^4 z~g_A^2~\epsilon^{\alpha\beta\gamma 3} ~\delta(x_3)~\left\{\bar\psi(z)\sigma^{\mu\nu}\left[\partial_{\mu}D_{\nu\alpha}(z-x)-\partial_\nu D_{\mu\alpha}(z-x) \right]\times \right. \nonumber\\
&&\qquad\qquad\qquad\qquad\qquad \left. G(z-w)\sigma^{\rho\tau}(i \partial_\beta) \left[\partial_\rho D_{\gamma\tau}(x-w)-\partial_\tau D_{\gamma\rho}(x-w)\right]\psi(w)\right\}\nonumber\\
&=&\frac{1}{2\pi}\bar u(p^{\prime})~ \delta(p^{\prime}-p)_{0,1,2}~\int d^4 k~ \left\{\sigma^{\mu\nu}\left[i(p^\prime-k)_\mu D_{\nu\alpha}(p^{\prime}-k)-i(p^\prime-k)_\nu D_{\mu\alpha}(p^\prime-k)\right]\times\right. \nonumber\\
&&\qquad\qquad\qquad\qquad\qquad \left. \epsilon^{\alpha\beta\gamma 3}~G(k)~\sigma^{\rho\tau}~(p-k)_\beta\left[i(p-k)_\rho D_{\gamma\tau}(p-k)-i(p-k)_\tau D_{\gamma\rho} (p-k)\right]\right\}u(p)
\end{eqnarray}
The second type scattering amplitude includes one CS vertex, one anomalous magnetic moment vertex and one normal electron-photon vertex $S_I^a S_I^b S_I^c$, which can be obtained from the following integral:
\begin{eqnarray}\label{A2}
M_{A2}&=&\int d^4x \int d^4 w\int d^4 z~g_A~\epsilon^{\alpha\beta\gamma 3}~\delta(x_3)\times\nonumber\\
& &~ \left\{\bar\psi(z)\sigma^{\mu\nu}\left[\partial_{\mu}D_{\nu\alpha}(z-x)-\partial_\nu D_{\mu\alpha}(z-x) \right]G(z-w)(ie\gamma^{\tau})   \partial_\beta D_{\gamma\tau}(x-w)\psi(w) \right\}\nonumber\\
&=&\frac{1}{2\pi}\bar u(p^{\prime})~ \delta(p^{\prime}-p)_{0,1,2}~\int d^4 k~ \left\{\sigma^{\mu\nu}\left[i(p^\prime-k)_\mu D_{\nu\alpha}(p^{\prime}-k)-i(p^\prime-k)_\nu D_{\mu\alpha}(p^\prime-k)\right]\times\right. \nonumber\\
&&\qquad\qquad\qquad\qquad\qquad \left. \epsilon^{\alpha\beta\gamma 3}~G(k)~\gamma^{\tau}~(p-k)_\beta  D_{\gamma\tau}(p-k) \right\}u(p)
\end{eqnarray}

From the expressions of Eq. \eqref{A1} and \eqref{A2}, one can find that $M_{A2}$ and $M_{A1}$, respectively, contain one and two more derivatives (momenta in the numerator) than $M$ in Eq. \eqref{scatteringAmplitude}.  With the same constraints from momentum conservation as in $M$, we find that both $M_{A1}\propto \int d^4 k~\frac{1}{k^2}$ and $M_{A2}\propto \int d^4 k~\frac{1}{k^3}$ diverge in the ultraviolet limit.  This is not a physical contradiction, however, because in reality both the anomalous magnetic moment term and the original action (Chern-Simons term) will have form factors that can relieve the divergence of the amplitude.  In fact, when integrating with frequency for a material, one has to choose a physical cutoff, e.g. the plasma frequency $\omega_p$.

\section{II. Theoretical framework for calculating atomic energy level shifts}

In this section, we outline the theoretical framework on how to calculate energy level shifts of an atom close to a material. Writing in a self-contained manner, we start from Lagrangian, and then calculate the two-photon exchange (second-order perturbation) result. 

We consider an atom or a molecular complex close to a material body. The total action can be written as $S=S_0+S_I$, where $S_0$ and $S_I$ represent free part and interaction part, respectively.

The free part  $S_0$ includes three parts, i.e.,
\begin{eqnarray}
 S_0=S_a+S_m+S_{em},
\end{eqnarray}
where $S_a$, $S_m$, and $S_{em}$ represent action for the atom, the material, and the electromagnetic field. We use the natural unit $\hbar=c=1$ in this note.
The interacting part includes the dipole interaction between electron and field:
\begin{eqnarray}
S_I=\int d^3x dt~\bar\psi_n ~ D_i B_i~\psi_m,
\end{eqnarray}
where $D_i$ stands for $i-$th component of magnetic moment operator $\vec D$, and the bound state $\psi_n(r)=\psi (\bold r)e^{-i \epsilon_n t}$  is a solution of the time dependent Dirac equation in an external potential $V(\vec r)$.  

The two-photon scattering matrix for $\psi_n$ is
\begin{eqnarray}
M(\bold r,\bold r^{\prime};t,t^\prime)
=-\frac{1}{2}\sum_m \langle n| D_i|m\rangle\langle m| D_j|n\rangle e^{i (\epsilon_n-\epsilon_m)(t-t^\prime)}\langle B_i(\bold r,t)B_j(\bold r^\prime,t^\prime)\rangle
\end{eqnarray} 
Here, due to fluctuation-dissipation theorem, $\langle B_i(\bold r,t)B_j(\bold r^\prime, t^\prime)\rangle=2\int_{-\infty}^{\infty}\frac{d\omega}{2\pi}~n(\omega, T)~ {\rm Im} \mathcal H_{ij}(\bold r,\bold r^\prime,\omega)~e^{-i\omega(t-t^\prime)}$, where $\mathcal H_{ij}$ is the magnetic Green's tensor, and $n(\omega, T)=(1-e^{-\hbar\omega/k_B T})^{-1}$ is the bosonic distribution function.  In the limit of zero temperature, $n(\omega,T=0)\rightarrow \Theta(\omega)$. 

For an atom located at $\bold r$, the energy shift (at zero temperature) is 
\begin{eqnarray}\label{deltaepsilonn}
\delta\epsilon_n=&&\mathcal P~\frac{i}{T}\int dt \int dt^\prime M(\bold r,\bold r^\prime, t, t^\prime)\nonumber\\
=&&-\frac{1}{2\pi} ~\mathcal P~\sum_m \int_{0}^\infty d\omega~ \frac{\langle n| D_i |m\rangle\langle m| D_j|n\rangle~ {\rm Im}~\mathcal H_{ij}(\bold r,\bold r^\prime,\omega)}{\omega-\epsilon_{nm}}.
\end{eqnarray}

By employing Dirac identity $\mathcal P[\frac{1}{x}]=\frac{1}{x+i\eta}+i\pi \delta(x)$, one can  separate the total energy shift into the off-resonant part $\delta \epsilon_n^{(1)}$ and the resonant part $\delta \epsilon_n^{(2)}$, i.e., $\delta\epsilon_n=\delta \epsilon_n^{(1)}+\delta \epsilon_n^{(2)}$,
where
\begin{eqnarray}
\delta \epsilon_n^{(1)}&=&-\frac{1}{2\pi}\sum_m \langle n| D_i |m\rangle\langle m| D_j|n\rangle~\int_0^\infty d\xi ~\frac{\epsilon_{mn} ~\mathcal H_{ij}(\bold r_0,\bold r_0;i\xi)}{\xi^2+\epsilon_{mn}^2}; \label{epsilon1} \\
\delta \epsilon_n^{(2)}&=&-\frac{1}{2}~\sum_m\langle n| D_i|m\rangle\langle m| D_j|n\rangle ~~{\rm Re}\left\{ \left[ \mathcal H_{ij}(\bold r,\bold r,\epsilon_n-\epsilon_m)\right]\right\}~\Theta(\epsilon_{nm}). \label{epsilon2}
\end{eqnarray}
Here, $\epsilon_{nm}=\epsilon_n-\epsilon_m$, and $\delta\epsilon_n^{(2)}$ only exists for excited state, i.e., $\epsilon_{nm}>0$. The resonant term can be usually ignored, because, in the expression of $\epsilon_n^{(2)}$,  the frequency of virtual photons has to be high enough to match the energy level spacing.  If we express the $|n\rangle$-state-specific polarizability as  $\beta_{ij}(\omega)=\sum_{m} \frac{2\epsilon_{mn}~\langle n | D_i|m\rangle\langle m| D_j|n\rangle}{\epsilon_{nm}^2-\omega^2}$.  Then, we can obtain the well-known formula \cite{Intravaia}
\begin{eqnarray}
\delta\epsilon_{n}^{(1)}&&=-\frac{1}{2\pi}\int_0^\infty d\xi~ \beta_{ij}(i\xi) ~\mathcal H_{ij}(\bold r_0,\bold r_0;i\xi)\nonumber\\
&&=-\frac{1}{2\pi} \int_0^{\infty} d\omega ~{\rm Im}\left\{\beta_{ij}(\omega)~\mathcal H_{ij}(\bold r_0,\bold r_0;\omega)\right\}.\label{polarizability}
\end{eqnarray}
\bigskip

\section{III. Quantum atmosphere of superconductors}

In this section, we calculate the atmosphere effect for normal superconductors and chiral superconductors. 

\subsection{1. Quantum atmosphere of normal superconductors}

The magnetic field Green's tensor can be written as a combination $\mathcal H_{ij}=\mathcal H_{ij}^{0}+\mathcal H_{ij}^s$. Here, $\mathcal H_{ij}^0$ stands for the Green's tensor in vacuum, whereas $\mathcal H_{ij}^s$ represents the Green's tensor contributed from the presence of the surface.  In the following, we only use $\mathcal H^{s}_{ij}$ in order to taking account into the surface effect. 
The surface magnetic Green's tensor can be generally written as
\begin{eqnarray}
\mathcal H(\bold r,\bold r^{\prime},\omega)=\frac{i}{2\pi}\int\,\frac{d k_x dk_y}{k_z}\,e^{ik_x(x-x^{\prime})}e^{ik_y(y-y^{\prime})}e^{ik_z(z+z^{\prime})}\left[r_{ss}M_{ss}+r_{pp}M_{pp}+r_{sp}M_{sp}+r_{ps}M_{ps}\right],
\end{eqnarray}
where $r_{ss}$, $r_{sp}$, $r_{ps}$, $r_{pp}$ represent the reflection coefficients for s (p)-polarized photons, and ${M}_{ss}, {M}_{pp}, {M}_{sp}$, and ${M}_{ps}$ are the corresponding Green's tensor matrices \cite{JCrosse}. Note that the magnetic Green's tensor $\mathcal H$ can be obtained from the electric one $\mathcal G$ by swapping the (s,p) sub-indices, i.e., $\mathcal H=\mathcal G(s\leftrightarrow p)$ \cite{Intravaia,chenkel}.  

For normal superconductors, $r_{ss}\rightarrow -1$ and $r_{pp}\rightarrow 1$, and $r_{sp}=r_{ps}=0$.
Substitute the Green's tensor into the expression \eqref{polarizability}, and one can obtain the  $n$th-level energy shift:

{\it Short range behavior}

\begin{eqnarray}\label{shortrb}
\delta \epsilon_n=\frac{1}{64\pi z^3}\langle n|\vec D\cdot \vec D+ D_z D_z |n \rangle
\end{eqnarray}
\\

{\it Long range behavior}
\\
\begin{eqnarray}\label{longrb}
\delta\epsilon_n=\frac{1}{64\pi z^4}\sum_m \frac{1}{\epsilon_{mn}} \langle n| D_i|m\rangle\langle m| D_i|n\rangle
\end{eqnarray}

Eqs \eqref{shortrb} and \eqref{longrb} have been also obtained in the reference \cite{haakh}.

\subsection{2. Quantum atmosphere of chiral superconductors}

In sharp contrast to the normal superconductors, time-reversal symmetry is broken in chiral superconductors.  Therefore, the chiral superconductors can support the nonzero cross-reflection coefficients, i.e., $r_{sp},~r_{ps}\neq 0$. Furthermore, the rotational symmetry at the surface leads to the equality $r_{sp}=r_{ps}$ \cite{jHam}. 
We calculate the spectra shift for an atom in the vicinity of a chiral superconductors,  and we show that the atom  ``sees" an effective long-range Zeeman field. 

The Green's tensor can have non-vanishing off-diagonal elements:
\begin{eqnarray}
\mathcal H_{12}(z,z; \omega)=-\mathcal H_{21}(z,z; \omega)=&&-\frac{e^{2 i\omega z}(1-2 i\omega z)i\omega}{4z^2}~ r_{sp}(\omega).
\end{eqnarray}
Notice that the Green's tensors $\mathcal H_{12}$ and $\mathcal H_{21}$ fulfills the relation $\mathcal H(\bold r, \bold r^\prime; \omega)\rightarrow \bold 0$ if $|\bold r-\bold r|\rightarrow \infty$, and the Schwarz reflection principle $\mathcal H^* (\bold r, \bold r^\prime; \omega)=\mathcal H(\bold r, \bold r^\prime; -\omega^*)$. However, since there is no time-reversal symmetry, the Lorentz's reciprocity principle is violated, i.e., $\mathcal H_{12}(\bold r,\bold r^\prime,\omega)\neq \mathcal H_{21}(\bold r^\prime,\bold r,\omega)$.

Substitute the magnetic Green's tensors into the expression \eqref{polarizability}, one can derive the energy level shift for chiral superconductors.

\begin{eqnarray}
\delta \epsilon_n &&=\frac{1}{2\pi}\sum_m \int_0^\infty d\omega~ \left(\frac{2\epsilon_{mn}}{\epsilon_{mn}^2-\omega^2}\right)~{\rm Im}\left\{\langle n| D_1|m\rangle\langle m| D_2|n\rangle \mathcal H_{12}(z,z;\omega)+\langle n| D_2|m\rangle\langle m| D_1|n\rangle \mathcal H_{21}(z,z;\omega)\right\}\nonumber\\
&&=\frac{g^2}{2\pi}\sum_m \int_0^\infty d\omega~ \left(\frac{2\epsilon_{mn}}{\epsilon_{mn}^2-\omega^2}\right)~{\rm Im}\left\{\langle n| S_1|m\rangle\langle m| S_2|n\rangle \mathcal H_{12}(z,z;\omega)+\langle n| S_2|m\rangle\langle m| S_1|n\rangle \mathcal H_{21}(z,z;\omega)\right\}
\end{eqnarray}
where we have replaced the dipole operator by $\vec D=g \vec S$ in the above formula, where $S_i$ represent Pauli matrices and $g$ is the magnetic moment of the electron. 

Let's make a very crude approximation, i.e., (similar approximation was also used by Bethe in the Lamb shift paper \cite{Bethe})
\begin{eqnarray}
\sum_m \int_0^\infty d\omega~\left\{\frac{2\epsilon_{mn}}{\epsilon_{mn}^2-\omega^2}~{\rm Im}\left\{...\right\} \right\}\approx 
\int_0^\infty d\omega~\left\{\frac{2\langle\epsilon_{mn}\rangle}{\langle\epsilon_{mn}\rangle^2-\omega^2}~\sum_m{\rm Im}\left\{...\right\}\right\},
\end{eqnarray}
where $\langle \epsilon_{mn}\rangle$ is the average energy level spacing. 

Applying this approximation, one can derive the Zeeman energy

\begin{eqnarray}\label{epsichiral}
\delta \epsilon_n\approx \frac{g^2}{2\pi}\frac{\langle n| S_z|n\rangle}{z^2}\int_0^\infty d\omega~ {\rm Re}~\left\{\frac{\langle \epsilon_{mn}\rangle e^{2i \omega z}\left[(1-2 i\omega z)i\omega\right]~r_{sp}(\omega)}{\langle \epsilon_{mn}\rangle^2-\omega^2} \right\}
\end{eqnarray}

There are different models for the cross reflection coefficients $r_{sp}(\omega)$ of chiral superconductors. Specific results may depend on specific models. But, as long as time-reversal symmetry is broken, one can expect such an effective Zeeman-energy shift.

In contrast to the normal superconductors where spin-up and spin-down electrons shift same amount energy, for chiral superconductors, however, spin-up and spin-down electrons shift differently. In other words, the electrons can feel an effective magnetic field $B_{eff}$ in the vicinity of the chiral superconductor.

In the following, we provide a real example to calculate the value of the effective magnetic field $B_{\text{eff}}$.  

The material strontium ruthenate,  $\rm Sr_2RuO_4$,  is thought to be a chiral superconductor for several decades \cite{kallin}. Despite enormous efforts, there are still a lot of controversial opinions on whether $\rm Sr_2RuO_4$ is a real chiral superconductor. One important signature for chiral superconductors is that they break time-reversal symmetry.  People have devoted substantial efforts to confirm the time-reversal symmetry breaking of this materials, e.g., using the Kerr effect \cite{kallin,jing}. 

Here, our quantum atmosphere method could provide a new way to see the symmetry breaking state of the chiral superconductors. Let's make an estimation on the size of the quantum atmosphere effect.  In the following, we use the model that is studied in this paper \cite{jing}, which can fit the experimental data therein very well.  The cross reflection coefficients can be modeled as
\begin{eqnarray}\label{rsp}
r_{sp}(\omega)=r_{pp}\times \frac{\omega_p}{\tilde n \omega^2 \tau}\frac{\Delta}{\epsilon_F},
\end{eqnarray}
where $\Delta$ represents superconducting gap, $\tilde n$ is the effective refractive index, $\omega_p$ is plasma frequency, and $\tau$ represents scattering time. If one substitute the expression \eqref{rsp} into eq. \eqref{epsichiral}, one can obtain (assume $r_{pp}\approx 1$)
\begin{eqnarray}
\delta \epsilon_n\approx \frac{g^2}{2\pi}\frac{\omega_p\Delta}{\tilde n\tau\epsilon_F}\frac{\langle n| S_z|n\rangle}{z^2}\int_0^\infty d\omega~ {\rm Re}\left\{\frac{\langle \epsilon_{mn}\rangle e^{2 i\omega z}(1-2i\omega z) i}{\omega(\langle \epsilon_{mn}\rangle^2-\omega^2)}\right\}
\end{eqnarray}

Here, we are interest in the short-range limit, i.e., $\omega z<<1$ and $\omega<<\langle \epsilon_{mn}\rangle $. In this case, the above equation can be simplified to 
\begin{eqnarray}
\delta \epsilon_n&&\approx \frac{g^2}{4}\frac{\omega_p\Delta}{\tilde n\tau\epsilon_F}\frac{\langle n| S_z|n\rangle}{z^2}\frac{1}{\langle \epsilon_{mn}\rangle}
=gB_{\text{eff}},
\end{eqnarray}
where the effective magnetic field is $B_{\text{eff}}=\frac{g}{4}\frac{\omega_p\Delta}{\tilde n\tau\epsilon_F}\frac{1}{z^2}\frac{1}{\langle \epsilon_{mn}\rangle}$. We can estimate the effective magnetic field by putting in real parameters as $z=10$ nm, $\langle \epsilon_{mn}\rangle=1$ eV, and $g=2\mu_B$, where $\mu_B$ is the Bohr magneton. Other quantities are chosen as the paper \cite{jing}. Then, one can obtain $B_{\text{eff}}\approx 10^{-11}$ Gauss.